**Comparative Study of RF Sputtered $Ba_2(BiNb)O_6$ and $Ba_2Bi_{1.4}Nb_{0.6}O_6$ thin films and Effect of Fluorine Doping in Their Photoelectrochemical (PEC) Performances.**

*Tilak Poudel, Baicheng Weng, Corey Grice, Yanfa Yan, Xunming Deng*

**Abstract**

Bismuth-based ternary oxide photocatalysts are of considerable interest in photoelectrochemical water splitting. Yet these oxides are highly stable in different environment, their relative inertness limits the available synthesis routes to obtain desired stoichiometry on the final product. This report describes a method to prepare barium-bismuth-niobate (specifically, $Ba_2(BiNb)O_6$ and $Ba_2Bi_{1.4}Nb_{0.6}O_6$) target by sintering a mixture of individual oxides and successfully fabricate desired barium-bismuth-niobate thin film on TCO substrate. The surface morphology, physical, and chemical properties of thin film were systematically investigated. Compositional uniformity was obtained by sputtering in oxygen/argon plasma environment and higher photocatalytic activities were observed after subsequent surface treatments. Sodium fluoride surface treatment further enhanced the photocurrent density and improved electrode's stability against corrosion. This work further suggests a viable approach to improve the PEC performance of sputtered barium bismuth niobate by modulating their fundamental energy states.

**Introduction**

Solar energy is sustainable and most promising energy source for the future thanks to its far more abundance than our projected energy needs [1]. Ever since the discovery of photoelectric effect, researchers and engineers have been working with the idea of converting light into electric power or to generate fuels such as hydrogen from water [2] [3]. First of all, extensive researches have been conducted to explore binary and ternary oxide based photoanodes such as $TiO_2$ [4] [5], $WO_3$ [6] [7] [8] $Fe2O3$ [9], $SbVO4$ [10], $BiVO4$ [11] [12] [13] [14] [15] for the use in photoelectrochemical (PEC) cells because of their relatively high stability in resisting oxidative photo corrosion and their low fabrication cost. Secondly, researches in materials engineering to develop new composite materials for monolithically integrated water splitting technologies is receiving exceptional attention [16] [17]. But this integrated technology [18] also demands adjustable optoelectronic properties, high control of doping levels, low photocorrosion and chemical corrosion. To address these issues for an efficient and reliable hydrogen production however, urgency in discovering novel oxide materials and exploring beyond the ternary compound is ever growing.

A novel synthesis procedure to sputter barium bismuth niobium oxide thin film using high purity homemade target is explained. The design principles used to identify this metal oxide composition and microstructure was based upon density functional theory [19] that predicts higher rates of photogenerated charge carriers and transport within the material due to its easily manipulative band edges position in semiconductor-liquid junction. In addition, $Ba_2BiNbO_6$ comprised of favorable electron bonding orbitals and crystal symmetry within the microstructure which leverages our research efforts to fabricate these materials in both bulk powder and thin film forms. A number of researches have been done in depositing $Ba_2(BiNb)O_6$ and $Ba_2Bi_{1.4}Nb_{0.6}O_6$ thin films using solution method [20] [21] [22] [23] but precursor oxides possess inherit inertness which limits the chemical synthesis route. After considering molecular details of its surface chemistry [19], we deposited Barium Bismuth Niobate double perovskite by magnetron sputtering and explored its surface properties.

This report outlines a method to achieve uniform coating of $Ba_2(BiNb)O_6$ and $Ba_2(Bi_{1.4}Nb_{0.6})O_6$ thin film over the large area. The surface morphology, bandgap modulation parameters, optical and electrochemical properties have been also studied. In this study, we report the synthesis of high-quality $Ba(BiNb)O_6$ and $Ba_2(Bi_{1.4}Nb_{0.6})O_6$ thin films for photo-absorber layer by reactive sputtering.

**Experimental Details**

1) **$Ba(BiNb)O_6$ and $Ba_2(Bi_{1.4}Nb_{0.6})O_6$ target synthesis: Preparation of precursor powder**

Precursor powder was synthesized by a high-temperature solid-state reaction method by mixing barium oxide (BaO), bismuth oxide ($Bi_2O_3$), and niobium oxide ($Nb_2O_5$) keeping the stoichiometric ratios of the metal elements Ba:Bi:Nb were nominally 2:1:1 (however an excess of 3-5 % atomic Bi was also included to account for losses from evaporation) for $Ba_2(BiNb)O_6$ powder and 2:1.4:0.6 for $Ba_2(Bi_{1.4}Nb_{0.6})O_6$ precursor powder. The mixtures were homogenized using a rolling mixer and as-obtained yellowish precursor powders was transferred to a fused-silica crucible which was placed into an ambient-atmosphere electric muffle furnace. A fused-silica plate was placed over the crucible to mitigate the loss of any volatile components (particularly Bi vapor) without creating a gas-tight seal, thus allowing excess oxygen to be present during the annealing process. The $Ba_2(BiNb)O_6$ assembly was brought up to approximately 820 $^o$C over a period of 4 hours and then held at this temperature for another 140 hours. And the $Ba_2(Bi_{1.4}Nb_{0.6})O_6$ assembly was sintered at 850 $^o$C for 120 hours. The furnaces were then deactivated and allowed to cool down naturally. The resulting powders were strong solid chunks and were dark brown in appearance. These chunks were then mechanically crushed and grinded thoroughly using agate mortar and pestle to convert into a fine microparticle.

Approximately 20 grams of each fine powders were used to fabricate $Ba(BiNb)O_6$ and $Ba_2(Bi_{1.4}Nb_{0.6})O_6$ sputtering targets. The powders were loaded into stainless-steel target cups with a 2″ diameter cavity and pressed at room temperature using hydraulic press with an applied force of 12 tons for 20 minutes to get the high-purity target ready for installation.

2) **Thin Film Deposition: Radio Frequency (RF) Magnetron Sputtering**

The high-purity homemade targets were separately loaded into a custom-built sputtering chamber at different times. Depositions were performed on FTO coated glass substrates with substrate temperatures ranging from ambient to 270$^o$C. The $Ba_2(Bi_{1.4}Nb_{0.6})O_6$ and $Ba_2(BiNb)O_6$ thin films were prepared by radio frequency sputtering in argon/oxygen plasma and in argon/helium plasma environment as well. Sputtering powers were typically ranging from 40 – 70 W, and chamber pressures of 10 mTorr sustained by supply mixture of either 1) argon and oxygen gases with the total flow rate of 30 sccm or 2) argon and helium gases with the total flow rate of 30 Sccm. The ratio of gases used was varied and the effect of helium and/or oxygen incorporation in resulting thin films was also studied.

We found that the under the reactive sputtering, $O_2$ partial pressure affected the sputtering yield of metal atoms. Lower $O_2$ partial pressure increases the sputtering yield of bismuth, however helium partial pressure resulted in lowering the sputtering yield of metals and produced thinner films. As the processing temperature has effects on stoichiometry of bismuth niobate ceramics [24], effect of various temperatures during and after the deposition was studied. The sputtered films reported here were post annealed at 550 ºC in argon environment for 60 mins unless otherwise stated.

### 3) Characterization and measurement

The atomic ratios of metals in the precursor powder were determined using energy dispersive x-ray spectroscopy with Rigaku Cu Kα radiation. The surface morphology and bulk elemental composition of the thin films were characterized using a Hitachi scanning electron microscope (SEM) with built-in energy dispersive spectroscopy (EDS) attachment. EDS measurements for elemental analysis were taken of regions approximately 500 μm x 500 μm in area. Film thicknesses were measured using a DEKTAK profilometer to determine the step height at two locations namely at a tape-masked center and holder frame-masked edges to observe the thickness variation across the film. Bulk crystalline structure of the thin films was characterized using a Rigaku X-ray diffractometer using coupled 2θ Bragg-Brentano mode and a copper X-ray source (Kα Cu =1.54 Å). Phase assignments were made based on the Joint Committee on Power Diffraction Standards (JCPDS) database.

The optical absorbance spectrum was measured by a PerkinElmer lambda 1050 UV-vis-NIR spectrophotometer. The transmittance and reflectance of the samples were measured by optical spectrometer using 300 nm – 1500 nm wavelength and the band gaps were calculated using the following relation.

$$\alpha = \frac{1}{t} \ln \left[ \frac{(1-R)^2}{T} \right] \qquad (1)$$

Where α is the absorption coefficient, t is thickness, and R and T are reflection and transmission respectively. Now, if we plot $(\alpha h\nu)^n$ vs. $h\nu$, the we can get a straight line, the intercept of which gives us the band-gap value.
n= 2 for direct; n=1/2 for indirect transition.

### 4) Photoelectrochemical (PEC) measurements

The photoelectrochemical properties were investigated using BBNO films as working electrode. BBNO electrodes were prepared by cutting large 4″×4″ thin films deposited on Tec-15 substrates into smaller 0.75″×1.5″ rectangular shapes. Thin film was then covered by a non-conducting epoxy resin from the edges to design an electrode, which has a smaller uncovered area in the center (typically less than 1 $cm^2$) for light exposure. The film on one side of these electrodes was then mechanically removed by gentle scratches exposing underlying conducting layer of the substrate and coated with thin indium metal layer for better electrical contact. The PEC characterization was carried out using Voltalab potentiostat, in a three-electrode configuration in a quartz-windowed cell partially filled with electrolyte solution. The recorded potential versus Ag/AgCl ($E_{Ag/AgCl}$) in this work was converted into potential against reversible hydrogen electrode (RHE) using the Nernst equation (2) given below:

$$E_{RHE} = E_{\frac{Ag}{AgCl}} + 0.059 \times pH + 0.1976\ V \qquad (2)$$

The system was purged with nitrogen for 30 mins in order to remove possible oxygen dissolved in the electrolyte. Photoelectrochemical response was recorded on both the forward bias and reverse bias potential under illumination. The illumination source was 300 W Xe lamp calibrated and equipped with AM 1.5 filter. The light intensity of 100 mW/cm$^2$ was adjusted and calibrated using silicon photodiode.

**Results & Discussion**

1. **Compositional Variation (edge-effect)**

The primary focus of the study was centered on obtaining uniform sputtering of target materials onto the substrate by controlling stoichiometric ratio of Bismuth to Niobium within the range of 1:1 to 1.4:1 using reactive sputtering. The as deposited thin films under pure argon environment and argon/helium environment have shown severe non-uniformity across the deposited thin film. The compositional constituents around the corner matched with the target composition but there has been consistently bismuth deficient (or niobium rich) at the center of both films as shown in the Figure 1 for Ba(BiNb)O$_6$ and Figure 2 for Ba$_2$(Bi$_{1.4}$Nb$_{0.6}$)O$_6$ thin films. This non-uniform and off-stoichiometric center of the thin films also display significantly poor PEC response in comparison to its corner.

This report describes for the first time that the discrepancy in chemical composition across these thin films can be mitigated by reactive sputtering. The amount of oxygen incorporation during thin film deposition has significant impact on its atomic composition indicating that oxygen plasma can alter the deposition rate of metal atoms. The atomic ratios of metals in the resulting thin films were determined by using energy dispersive x-ray spectroscopy. EDS results recorded from 500μm × 500 μm area of room temperature deposition with the standard deviation of 10 measurements are presented. In the following discussion, we will concentrate on the optoelectronic properties of representative Ba$_2$(BiNb)O$_6$ and Ba$_2$(Bi$_{1.4}$Nb$_{0.6}$)O$_6$ thin films obtained by reactive sputtering using argon/oxygen partial pressure. EDS measurements from multiple points around center and corner showed compositional uniformity in as-deposited thin films. Better performing amples are usually 500-700 nm thick.

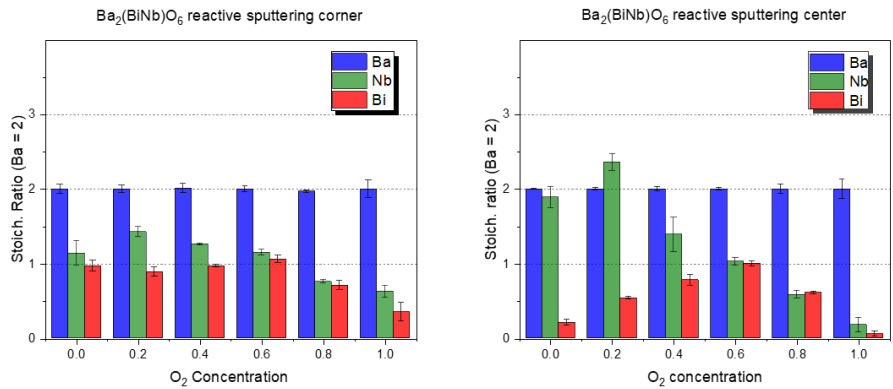

*Figure 1: elemental composition of Ba$_2$(BiNb)O$_6$ sputtered thin films on tec-15 substrate a) at corners b) at centers. Oxygen partial pressure of 0.6 produced near uniform thin films.*

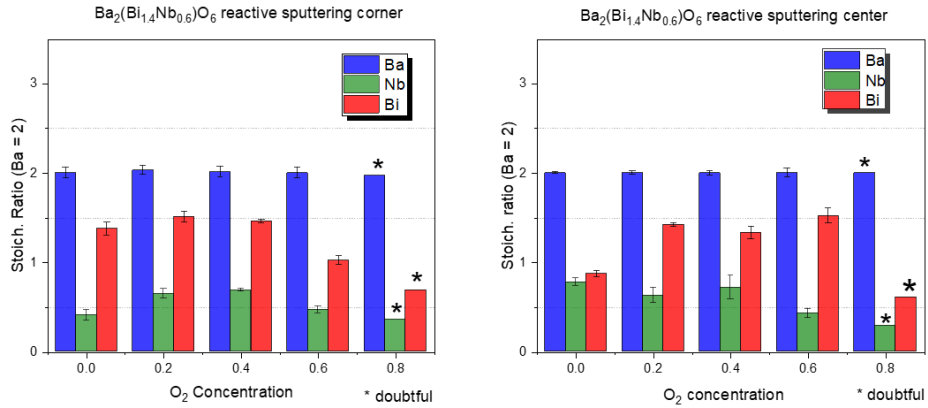

*Figure 2: Stoichiometry of metals in $Ba_2Bi_{1.4}O_6$ thin film deposited under varying oxygen partial pressure. Oxygen partial pressure of 0.2-0.4 produced near uniform desired thin films.*

Figure 3 below shows scanning electron microscopy images of argon annealed $Ba_2BiNbO_6$ and $Ba_2(Bi_{1.4}Nb_{0.6})O_6$ thin films. The grains of $Ba_2(Bi_{1.4}Nb_{0.6})O_6$ films are slightly bigger than the $Ba_2BiNbO_6$ thin films yet the grain boundaries are not clearly definite in either cases. The main reason might be the presence of residual stress in the bulk, which leads to increase microcracks on the surface with further increase in annealing temperature.

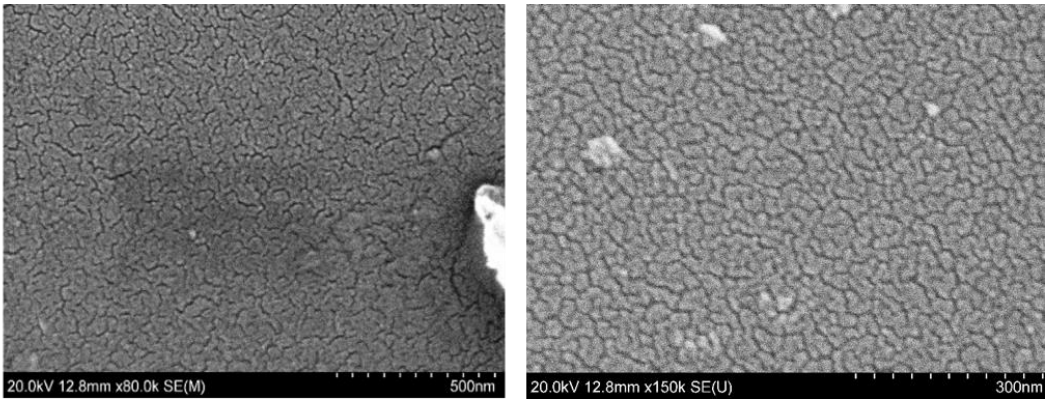

*Figure 3: Reactive rf Sputtered a) $Ba(BiNb)O_6$ and b) $Ba_2(Bi_{1.4}Nb_{0.6})O_6$ thin films after argon annealing at 550 ºC for 30 minutes.*

## Optical Properties

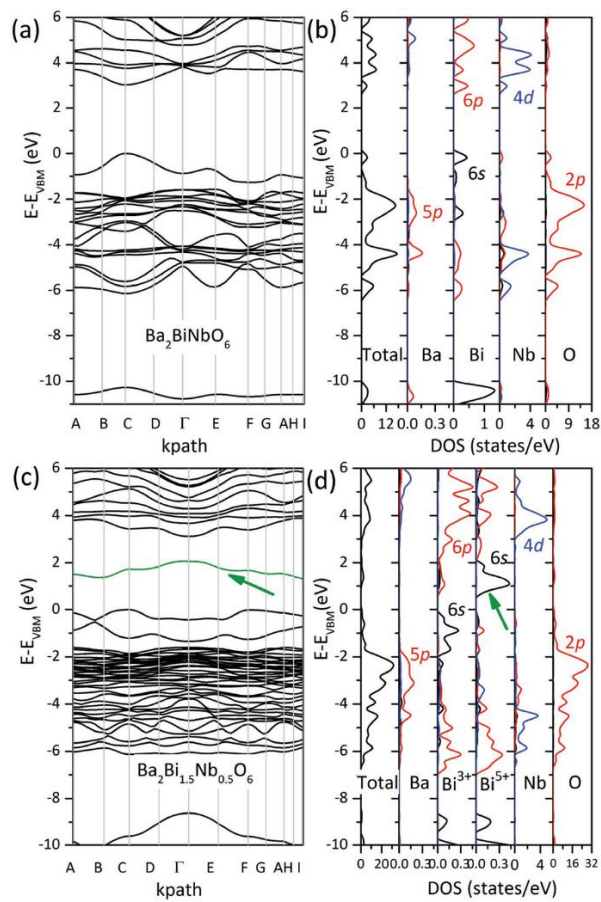

*Figure 4: a,c) Band structures and b,d) total and partial DOS of $Ba_2BiNbO_6$ and $Ba_2Bi_{1.5}Nb_{0.5}O_6$. (adapted from Baicheng et al. 2017)*

Contrary to the recent reports [25] [22] on solution-based $Ba_2(BiNb)O_6$ and $Ba_2(Bi_{1.4}Nb_{0.6})O_6$ thin films, we found a little higher direct band gap of 3.24 eV and 2.63 eV respectively for rf sputtered films as shown in Figure 5. The reduction in band gap for bismuth rich thin films is still concurrent with our previous work on density function theory [25] calculation, which predicted lowering of conduction band minimum due to bismuth 6s lone pair of electrons as films become more bi-rich. But the report was not clear about whether the transition is direct or indirect. The measurements in Figure 5 showed remarkable direct transition. The indirect band gap of 2.04 eV for $Ba_2(BiNb)O_6$ and 1.48 eV for $Ba_2(Bi_{1.4}Nb_{0.6})O_6$ was also calculated from reflection and transmission data as shown in supplementary information, but it's doubtful since the film thickness was not large enough to produce reliable absorption data.

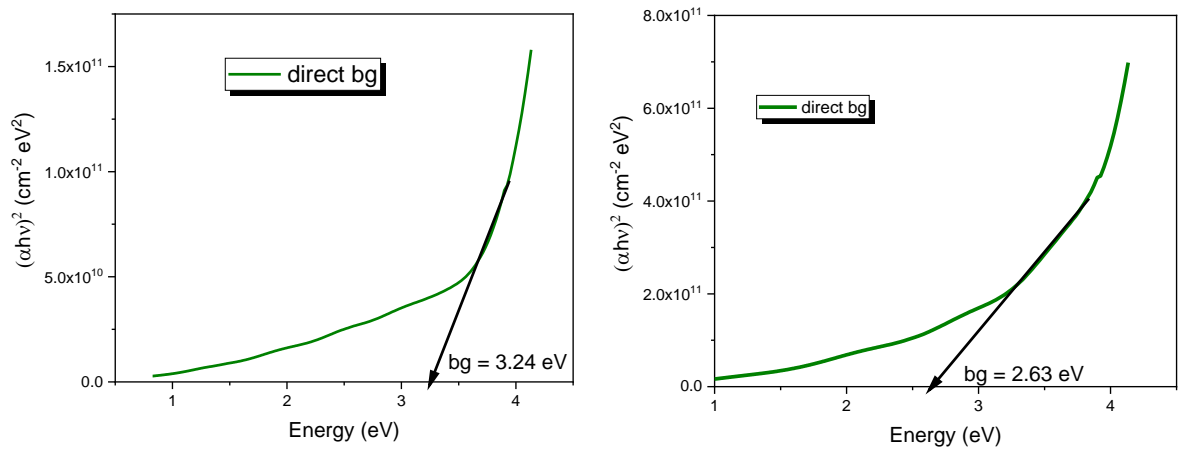

*Figure 5: Tauc Plot for direct band gap estimation of rf sputtered a) $Ba_2(BiNb)O_6$ thin films b) $Ba_2(Bi_{1.4}Nb_{0.6})O_6$ thin films.*

The experiments showed that band gap and light absorption of barium bismuth niobate can be manipulated with niobium concentration in the bulk. With lower niobium doping, the $Ba_2(Bi_{1.4}Nb_{0.6})O_6$ sputtered thin films display lower band gap of 2.63 eV as shown in Figure 5, which would correspond to the absorption of 470 nm and below. The Bi atoms occupying Nb sites have $Bi^{5+}$ oxidation states [26], the low energy-lying unoccupied $Bi^{5+}$ 6s orbital introduces a new conduction band edge below the conduction band edge of $Ba_2(BiNb)O_6$, resulting in a much-reduced bandgap [19] for bismuth rich thin films. Mixing $s^2$ $Bi^{3+}$ cations with $d^0$ cations $Nb^{5+}$ creates a coupling between the $s$ and the O $2p$ that forces an upward dispersion of the valence band [27], and since $Nb^{5+}$ is a donor ion in the crystal lattice it increases electrical conductivity by involving charge compensation by native vacancies [28].

**Photoelectrochemical (PEC) response**

1) **Edge-effect PEC comparison between $Ba_2(BiNb)O_6$ and $Ba_2(Bi_{1.4}Nb_{0.6})O_6$ thin films**

Both of the $Ba_2(BiNb)O_6$ and $Ba_2(Bi_{1.4}Nb_{0.6})O_6$ thin film displays visibly distinguished white center and dark brown edges (see supplementary information). The electrodes were designed from both the center part (white region) and edges (dark brown color) after annealing at 550 °C and compared their PEC performance as well. EDS confirmed the centers of the both films were bismuth poor (Figure 1 and Figure 2).

Preliminary photo response of electrodes was measured by illuminating films with a 300W Xe arc lamp while immersed in 1 M $Na_2SO_3$ + O.5 M $K_2HPO_4$ (pH = 7.0) electrolyte solution (after purging with nitrogen for over 30 minutes) as shown in Figure 6. It was found that the bi-poor centers in Figure 6 ( deposited under no oxygen partial pressure) yielded almost no photocurrent, most likely due to poor electrical contact between individual grains and possibly high defect concentrations at the grain boundaries. The dark brown stoichiometric edges, however, displayed small photo current in forward bias region.

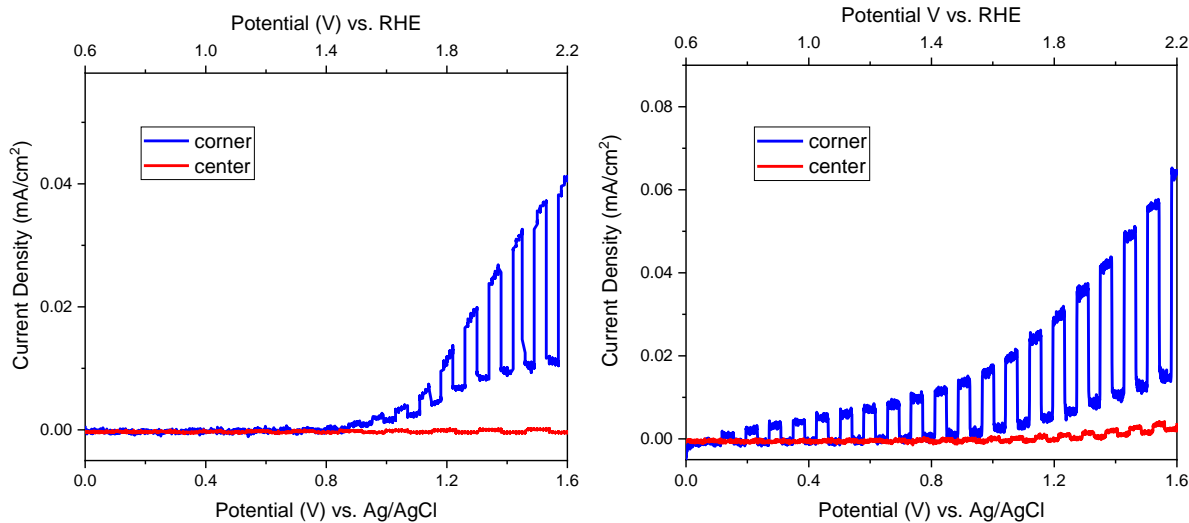

*Figure 6: PEC response variation between corners and centers of a)$Ba_2(BiNb)O_6$ b)$Ba_2(Bi_{1.4}Nb_{0.6})O_6$ thin films in pH 7 electrolyte solution. Centers of the deposited films are consistently bismuth poor. The samples were annealed at 550 °C for 30 minutes.*

The issue of non-uniformity in argon deposited thin films can be resolved by adding oxygen partial pressure during deposition. We were able to fabricate uniform $Ba_2(BiNb)O_6$ thin film under 0.6 oxygen partial pressure and uniform $Ba_2(Bi_{1.4}Nb_{0.6})O_6$ under 0.2 oxygen partial pressure during sputtering. It appears that the scattering of heavier atoms depends on amount of oxygen ions present along the path during deposition. EDS study of elemental composition across the film was consistent (Figure 1 and Figure 2), we designed the electrode from central region and edges, both of which displayed similar photocurrent response as shown in Figure 7. But the overall photo-response of $Ba_2(Bi_{1.4}Nb_{0.6})O_6$ is greater and more consistent than $Ba_2(BiNb)O_6$ thin films performances.

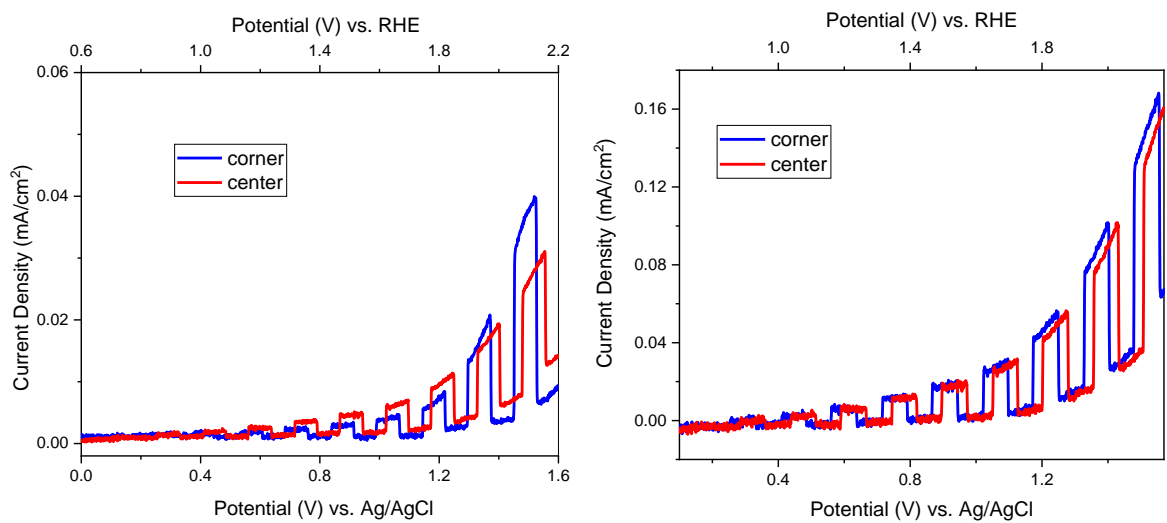

*Figure 7: corner vs. center pec response of argon/oxygen environment sputtered thin films a) $Ba_2BiNbO_6$ b) $Ba_2Bi_{1.4}Nb_{0.6}O_6$ thin films in pH 7.0 electrolyte solution.*

## 2) Ambipolar PEC of oxygenated $Ba_2(Bi_{1.4}Nb_{0.6})O_6$ thin films

Investigations showed that the oxygenated $Ba_2Bi_{1.4}Nb_{0.6}O_6$ thin film displaced both the n-type and p-type characteristics in aqueous media as shown in Figure 8. The ambipolar behavior matches with the reports on other double perovskites containing both $Bi^{3+}$ and $Bi^{5+}$ oxidation states along with its tantalum, lead, and antimony derivates have been demonstrated p-type characteristics as well [22] [29] [30]. Oxygenated $Ba_2Bi_{1.4}Nb_{0.6}O_6$ film appears to have a bulk electrical resistance in the order of $10^7$ Ω cm, comparable to solution-based $Ba_2Bi_{1.4}Nb_{0.6}O_6$ from our previous work [19]. While the films are highly resistive, its ambipolar photocurrent response can be explained on the basis of Fermi level shifting. For the $Ba_2Bi_{1.4}Nb_{0.6}O_6$ film (Bi-rich), the surface Fermi level lies further away from the conduction band minimum than in the $Ba_2BiNbO_6$ film since Bi:Nb =1:1 thin film has its fermi level closer to the conduction band. It may give the explanation for lower band gap for bi-rich thin films as shown in Figure 5.

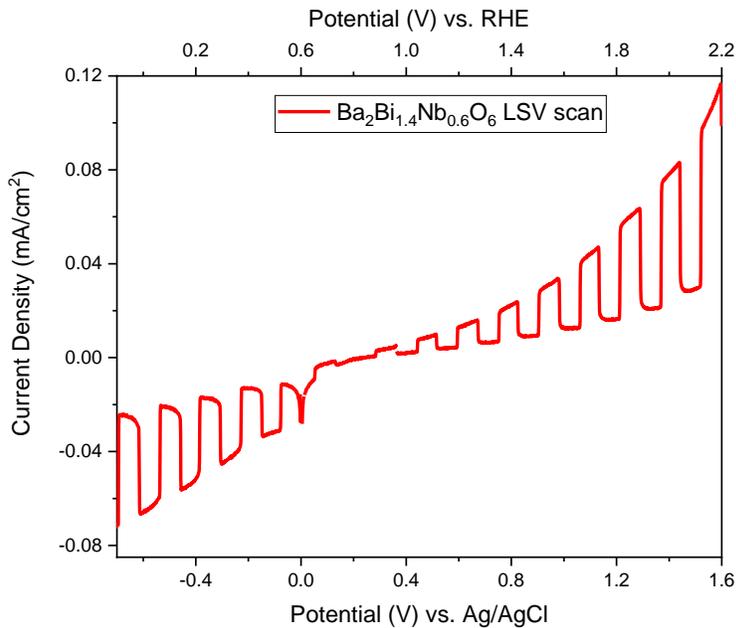

Figure 8: Linear sweep voltammetry graphs of as sputtered $Ba_2Bi_{1.4}Nb_{0.6}O_6$ thin films in pH 7.0 electrolyte solution. The argon annealed films exhibit small ambipolar current density.

It was also discovered that the ambipolar films are stable with the application of positive bias up to 1.23 volt versus Ag/AgCl reference electrode, while all films displayed significant degradation (color fading to brownish and eventual delamination) upon the application of negative bias beginning around -0.6 volt versus Ag/AgCl reference electrode. We also observed thin film corrosion in alkaline electrolyte solution as shown in Figure 9b below. Electrode degradation could be resulted from the reduction of metal cations (Bi and/or Nb). Nevertheless, for photoanode applications, the negative bias potentials are unlikely to be experienced and should not possess significant threat.

3) **Effect of Sodium Fluoride surface treatment**

The low (in the order of few microamperes per square centimeter) yet ambipolar current density is not in very much useful for the water oxidation or reduction purposes since only one type of reaction can occur on the surface of working electrode. Thus, changing the carrier type is especially important since photoanodes operate most effectively when they have n-type characteristics.
An effective way of increasing reactivity of these highly resistive oxides materials is by doping. Investigations on manipulating the carrier type and concentration by incorporating dopant atoms such as Cl, F, and Mo was studied by spraying 0.001 M $BaCl_2$ solution, 0.005 M NaF solution, and thermal evaporation of 2 nm of $MoO_3$ respectively and consequent annealing in argon environment at 550 ºC for 30 minutes. Our results indicate that $BaCl_2$ treatment didn't show significant improvement on photocurrent density and $MoO_3$ surface treatment even diminished current density (supplementary information). However, NaF treated samples showed better photo response at higher bias and induced cathodic shift in onset potential (Figure 9), maybe because fluorine penetrates the bulk and replaces oxygen from its lattice during surface treatments owing to its relatively similar atomic size and high reactivity.

Sodium fluoride treated samples showed improved photo-response under all acidic, neutral, and alkaline electrolyte solution (Figure 9 and supplementary information). In $Ba_2(BiNb)O_6$ thin films, NaF treatment enhanced photocurrent density significantly (up to 0.1 mA/cm$^2$ at 1.23 V against Ag/AgCl) and in $Ba_2Bi_{1.4}Nb_{0.6}O_6$ thin films, NaF treatment can alters not only its carrier type but also the photocurrent density as shown in Figure 9. Not only switching the electrodes into more n-type, NaF treatment also increase the photo current density by the factor of 2 or more at higher bias region (beyond 0.8 volt against Ag/AgCl). In addition, treatment of NaF results into reducing dark current in $Ba_2Bi_{1.4}Nb_{0.6}O_6$ thin films. High dark current is typically responsible for photo-degradation of submerged electrodes. Finally, all films appear to be stable in the aqueous electrolytes (with applied bias under 1.0 V vs. Ag/AgCl) for several hours (>7 hrs) without any visible signs of degradation.

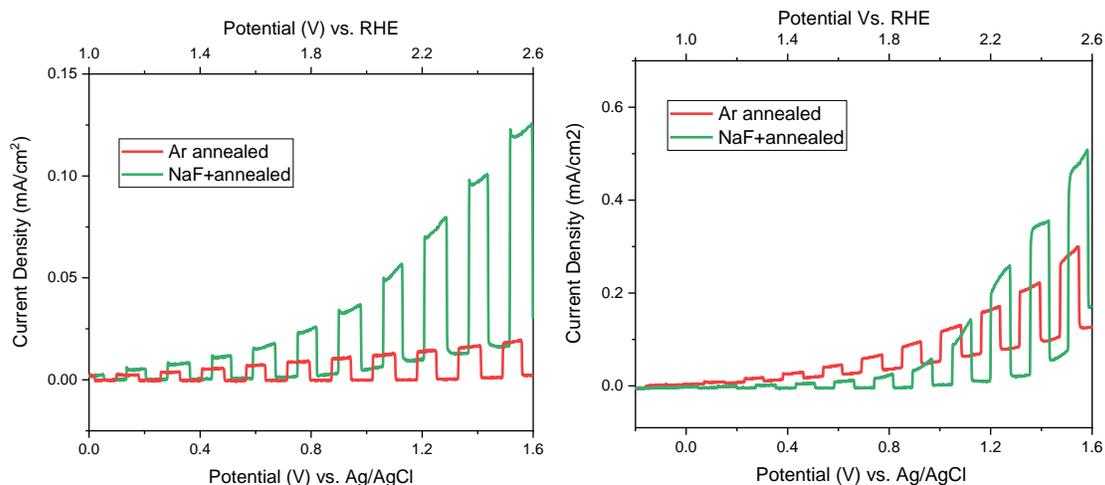

*Figure 9: PEC response comparison between before and after NaF surface treatments of oxygenated a) $Ba_2(BiNb)O_6$ thin films and b) $Ba_2Bi_{1.4}Nb_{0.6}O_6$ thin films in 0.5 M KOH electrolyte solution.*

The photoelectrochemical (PEC) performance of the $Ba_2Bi_{1.4}Nb_{0.6}O_6$ thin films after NaF surface treatment is among the best reported in the literature for single junction sputtered $Ba_2Bi_{1.4}Nb_{0.6}O_6$ thin film.

**Effects of Annealing Temperature**

Lower processing temperature does not necessary remove all the defects and recombination centers in the film, which suggests higher processing temperature. Normally, the grain sizes increase upon annealing. Films annealed at 550 ºC in argon environment showed better performance even without the fully-grown grains on the surface (as seen in SEM images Figure 3). Nonetheless, annealing at 600 ºC and above gives $Ba_2(BiNb)O_6$ thin films exhibit strain developed along the substrate and $Ba_2Bi_{1.4}Nb_{0.6}O_6$ thin films suffer delamination. This phenomenon can be ascribed to the thermal expansion mismatch between $Bi_2O_3$ and $Nb_2O_3$, which develops the tensile stresses across the thin film [31].

**Conclusion**

The successful synthesis of $Ba_2(BiNb)O_6$ and $Ba_2Bi_{1.4}Nb_{0.6}O_6$ thin films using reactive RF sputtering is reported for the first time. By incorporating oxygen in the plasma environment, we were able to eliminate center and corner effects to obtain stoichiometrically uniform thin films of various thickness ranging from 500 nm – 700 nm. In addition, we studied their optical and photoelectrochemical properties with possible surface modifications. It appears that the band gap of these materials depends on doping level of niobium in the crystal lattice. Increasing bismuth concentration from 1 to 1.4 reduces the band gap by 0.61 eV. Various surface modification designs were implemented, and we found that the NaF treatment can change the ambipolar $Ba_2Bi_{1.4}Nb_{0.6}O_6$ thin films into more n-type materials. We have also found that sodium fluoride treatment on both films significantly suppress the dark photocurrent and hence enhance the overall photovoltage.

# Bibliography

[1] J. Luo, J.-H. Im, M. T. Mayer, M. Schreier, M. K. Nazeeruddin, N.-G. Park, S. D. Tilley, H. J. Fan and M. Gratzel, "Water photolysis at 12.3% efficiency via perovskite photovoltaics and Earth-abundant catalysts," *Science,* vol. 345, no. 6204, pp. 1593-1596, 2014.

[2] M. Gratzel, "Photoelectrochemical cells," *Nature,* vol. 414, pp. 338-344, 2001.

[3] S. Y. Reece, J. A. Hamel, K. Sung, T. D. Jarvi, A. J. Esswein, J. J. H. Pijpers and D. G. Nocera, "Wireless Solar Water Splitting Using Silicon-Based Semiconductors and Earth-Abundant Catalysts," *Science,* vol. 334, pp. 645-648, 2011.

[4] L. Chen, M. E. Graham, G. Li and K. A. Gray, "Fabricating highly active mixed phase TiO2 photocatalysts by reactive DC magnetron sputter deposition," *Thin Solid Films,* pp. 1176-1181, 2006.

[5] G. Wang, H. Wang, Y. Ling, Y. Tang, X. Yang, R. C. Fitzmorris, C. Wang, J. Z. Zhang and Y. Li, "Hydrogen-Treated TiO2 Nanowire Arrays for Photoelectrochemical Water Splitting," *Nano Lett.,* pp. 3026-3033, 2011.

[6] J. Cen, Q. Wu, D. Yan, W. Zhang, Y. Zhao, X. Tongb, M. Liu and A. Orlov, "New aspects of improving the performance of WO3 thin films for photoelectrochemical water splitting by tuning the ultrathin depletion region," *Royal Society of Chemistry,* pp. 899-905, 2019.

[7] X. Liu, F. Wanga and Q. Wanga, "Nanostructure-based WO3 photoanodes for photoelectrochemical water splitting," *Physical Chemistry Chemical Physics,* pp. 7894-7911, 2012.

[8] S. S. Kalanur, I.-H. Yoo, I. S. Cho and H. Seo, "Niobium incorporated WO3 nanotriangles: Band edge insights and improved photoelectrochemical water splitting activity," *Ceramics International,* vol. 45, no. 7, pp. 8157-8165, 2019.

[9] D. K. Sivula and P. D. M. G. Florian Le Formal, "Solar Water Splitting: Progress Using Hematite (α-Fe2O3) Photoelectrodes," *ChemSusChem,* 2011.

[10] A. R. L. Canovas, J. F. Garcia and S. Hansen, "Structural flexibility in SbVO4," *Catalysis Today,* vol. 158, pp. 156-161, 2010.

[11] J. K. Cooper, S. Gul, F. M. Toma, L. Chen, Y.-S. Liu, J. Guo, J. W. Ager, J. Yano and I. D. Sharp, "Indirect Bandgap and Optical Properties of Monoclinic Bismuth Vanadate," *The Journal of Physical Chemistry,* pp. 2969-2974, 2015.

[12] Y. Qiu, W. Liu, W. Chen, W. Chen, G. Zhou, P.-C. Hsu, R. Zhang, Z. Liang, S. Fan, Y. Zhang and Y. Cui, "Efficient solar-driven water splitting by nanocone BiVO4-perovskite tandem cells," *Science Advances,* vol. 2, 2016.

[13] P. M. Rao, L. Cai, C. Liu, I. S. Cho, C. H. Lee, J. M. Weisse, P. Yang and X. Zheng, "Simultaneously efficient light absorption and charge separation in WO3/BiVO4 core/shell nanowire photoanode for photoelectrochemical water oxidation.," *Nano Letters,* pp. 1099-1105, 2014.

[14] T. W. Kim and K.-S. Choi, "Nanoporous BiVO4 Photoanodes with Dual-Layer Oxygen Evolution Catalysts for Solar Water Splitting," *Science,* vol. 343, 2014.

[15] Y. Park, K. J. McDonald and K.-S. Choi, "Progress in bismuth vanadate photoandoes for use in solar water oxidation," *The Royal Society of Chemistry,* vol. 42, pp. 2321-2337, 2013.

[16] J. Shah, S. Jain, A. Shukla, R. Gupta and R. K. Kotnala, "A Facile non-photocatalytic technique for hydrogen gas production by hydroelectric cell," *International journal of hydrogen energy,* 2017.

[17] M. M. May, H.-J. Lewerenz, D. Lackner, F. Dimroth and T. Hannappel, "Efficient direct solar-to-hydrogen conversion by in situ interface transformation of a tandem structure," *Nature Communications,* 2015.

[18] M. G. Walter, E. L. Warren, J. R. McKone, S. W. Boettcher, Q. Mi, E. A. Santori and N. S. Lewis, "Solar Water Splitting Cells," *Chem. Rev. ,* pp. 6446-6473, 2010.

[19] B. Weng, Z. Xiao, W. Meng, C. R. Grice, T. Poudel, X. Deng and Y. Yan, "Bandgap Engineering of Barium Bismuth Niobate Double Perovskite for Photoelectrochemical Water Oxidation," *Advanced Energy Materials,* vol. 7, p. 1602260, 2017.

[20] C. C. Tan, A. Feteira and D. C. Sinclair, "Ba2Bi1.4Nb0.6O6 Nonferroelectric, High Permittivity Oxide," *Chemistry of Materials,* pp. 2247-2249, 2012.

[21] N. Panda, B. N. Parida, R. Padhee and R. N. Choudhary, "Structural, dielectric and electric properties of the BaBiNbO6 double perovskite," *Journal of Materials Science Materials in Electronics,* 2015.

[22] R. V. K. Mangalam, P. Mandal, E. Suard and A. Sundaresan, "Ferroelectricity in Ordered Perovskite BaBi0.53+(Bi0.25+Nb0.35+)O3 with Bi3+:6s2 Lone Pair at the B-site," *Chemistry of Materials,* pp. 4114-4116, 2007.

[23] S. P. Gaikwad, H. S. Potdar, V. Samuel and V. Ravi, "Co-precipitation method for the preparation of fine ferroelectric BaBi2Nb2O9," *Ceramics International,* pp. 379-381, 2005.

[24] C.Karthik and K.B.R.Varma, "Influence of vanadium doping on the processing temperature and dielectric properties of barium bismuth niobate ceramics," *Materials Science and Engineering: B,* vol. 129, no. 1-3, pp. 245-250, 2006.

[25] B. Weng, C. R. Grice, J. Ge, T. Poudel, X. Deng and Y. Yan, "Barium Bismuth Niobate Double Perovskite/Tungsten Oxide Nanosheet Photoanode for High-Performance Photoelectrochemical Water Splitting," *Advanced Energy Materials,* vol. 8, p. 1701655, 2018.

[26] W.-J. Yin, B. Weng, J. Ge4, Q. Sun, Z. Li and Y. Yan, "Oxide Perovskites, Double Perovskites and Derivatives for Electrocatalysis, Photocatalysis, and Photovoltaics," *Energy & Environmental Science,* vol. 12, pp. 442-462, 2019.

[27] K. Sivula and R. v. d. Krol, "Semiconducting materials for photoelectrochemical energy conversion," *Nature Reviews Materials,* p. 15010, 2016.

[28] L. R. Sheppard, A. J. Atanacio, T. Bak, J. Nowotny and K. E. Prince, "Bulk Diffusion of Niobium in Single-Crystal Titanium Dioxide," *Journal of Physical Chemistry B,* vol. 111, pp. 8126-8130, 2007.

[29] J. Ge, Yin, Wan-Jian and Y. Yan, "Solution-Processed Nb-Substituted BaBiO3 Double Perovskite Thin Films for Photoelectrochemical Water Reduction," *Chemistry of Materials,* vol. 30, no. 3, pp. 1017-1031, 2018.

[30] S.-H. Lee, J.-H. Sohn, J.-H. Lee and S.-H. Cho, "Dielectric loss anomaly of BaBiO3," *Journal of Applied Physics,* vol. 86, pp. 6351-6354, 1999.

[31] W. Mielcarek and K. Prociow, "BaBiO2.77 as a promoter of the varistor property in zinc oxide ceramics," *Journal of the European Ceramic Society,* vol. 21, no. 6, pp. 711-717, 2001.